Gate-modulated conductance of few-layer WSe$_2$ field-effect transistors in the subgap regime: Schottky barrier transistor and subgap impurity states

Junjie Wang[1], Daniel Rhodes[2], Simin Feng[1], Minh An T. Nguyen[3], K. Watanabe[4], T. Taniguchi[4], Thomas E. Mallouk [1,3,5], Mauricio Terrones[1,3,6,7], Luis Balicas[2], J. Zhu[1,7,a)]

[1]Department of Physics, The Pennsylvania State University, University Park, PA, 16802, USA
[2]National High Magnetic Field Lab, Florida State University, Tallahassee, FL, 32310, USA
[3]Department of Chemistry, The Pennsylvania State University, University Park, PA, 16802, USA
[4]National Institute for Materials Science, 1-1 Namiki, Tsukuba, 305-0044, Japan
[5]Department of Biochemistry and Molecular Biology, The Pennsylvania State University, University Park, PA, 16802, USA
[6]Department of Materials Science and Engineering, The Pennsylvania State University, University Park, PA, 16802, USA
[7]Center for 2-Dimensional and Layered Materials, The Pennsylvania State University, University Park, PA 16802, USA

____________________________

a) Electronic mail: jzhu@phys.psu.edu

Two key subjects stand out in the pursuit of semiconductor research: material quality and contact technology. The fledging field of atomically thin transition metal dichalcogenides (TMDCs) faces a number of challenges in both efforts. This work attempts to establish a connection between the two by examining the gate-dependent conductance of few-layer (1-5L) WSe$_2$ field effect devices. Measurements and modeling of the subgap regime reveal Schottky barrier transistor behavior. We show that transmission through the contact barrier is dominated by thermionic field emission (TFE) at room temperature, despite the lack of intentional doping. The TFE process arises due to a large number of subgap impurity states, the presence of which also leads to high mobility edge carrier densities. The density of states of such impurity states is self-consistently determined to be approximately 1-2×10$^{13}$ /cm$^2$/eV in our devices. We demonstrate that substrate is unlikely to be a major source of the impurity states and suspect that lattice defects within the material itself are primarily responsible. Our experiments provide key information to advance the quality and understanding of TMDC materials and electrical devices.

Atomically thin transition metal dichalcogenides (TMDCs) MX$_2$ (M=Mo, W; X=S, Se, Te) are a new class of two-dimensional semiconductors with attractive electronic, optical, and valleytronic properties and application potential in the emerging area of 2D nanoelectronics.[1] While the syntheses of a large variety of binary compounds, alloys and vertical and lateral junctions are rapidly progressing[2-6] and many device concepts are being evaluated,[7-12] fundamental knowledge of their intrinsic electronic properties is fairly limited, partly due to the challenge of making ohmic contacts to thin sheets, a problem inherent to semiconductors with a sizable band gap. Recent studies suggest that contact resistance plays a dominant role in the field effect of TMDC transistors.[13] The impact of the contact metal work function was studied in several materials.[14-17] In MoS$_2$, for example, contact metals with diverse working functions ranging from 3.5 to 5.9 eV appear to all lie close to the conduction band,[14] thus suggesting Fermi level pinning by surface states.[18] The origin of this behavior needs to be understood before p-type devices can be made. Doping the contact region chemically or using electrolyte is shown to help, although a recipe compatible with large-scale device processes has yet to be developed.[19-21] In addition to contact challenges, the carrier mobility $\mu$ in TMDC materials is relatively low compared to conventional 2D systems. For example, the low-temperature field effect mobility $\mu_{FE}$ is below 1000 cm$^2$/Vs even in current high-quality monolayer MoS$_2$.[22, 23] Further improving the quality of TMDC materials can greatly facilitate the exploration of fundamental phenomena in these fascinating two-dimensional systems.[24]

In this letter, we focus on the measurement and understanding of the gate-dependent conductance $G(V_{bg})$ of few-layer (1-5L) WSe$_2$ field effect transistors to illuminate the issues of contact and disorder mentioned above. Applying the gate voltage in pulse eliminates hysteresis in $G(V_{bg})$, which allows us to probe the intrinsic charging of the WSe$_2$ sheet, free of the influence of the dielectric trap states. Devices constructed on different substrates are studied to examine the effect of the substrate/WSe$_2$ interface. Our results show that below the mobility edge, $G(V_{bg})$ is dominated by gate-modulated transmission through the Schottky barrier contacts and the primary transmission mechanism is *thermionic field emission*. We establish a quantitative connection between the observed sub-threshold swing and the density of states (DoS) of the subgap impurity states, the latter is estimated to be ~1-2×10$^{13}$/cm$^2$/eV from evaluating many devices. We also discuss possible sources of the impurity states. These results offer insights to the understanding of transport measurements in TMDC devices as well as provide key input to further improving their performances.

Chemical vapor transport (CVT) methods are used to grow bulk crystals of WSe$_2$, from which we mechanically extract few-layer sheets (See S1, Ref. 51). Figure 1(a) shows a high-angle annular dark field (HAADF) transmission electron microscopy (TEM) image of a WSe$_2$ crystal along the [110] plane of both W and Se atoms (grown by the first method). A clear hexagon lattice confirms the 2H phase of the crystal and its high crystallinity. Energy-dispersive x-ray (EDX) spectrum yields an atomic ratio of 33% W and 67% Se (±1%), confirming its stoichiometry (data not shown). Figure 1(b) shows an x-ray diffraction (XRD) pattern of a WSe$_2$ crystal grown by the second method. We obtain in-plane and out-of-plane lattice constants $a$=3.28 Å and $c$=12.95 Å respectively from the XRD data, which agrees very well with prior reports on the crystal structure of 2H WSe$_2$.[25] Figure 1(c) plots the temperature-dependent micro-photoluminescence (μ-PL) spectra of a monolayer WSe$_2$ sheet. Lorentz fitting reveals two peaks, which we attribute to the A-exciton and the trion respectively following the literature.[26] The temperature-dependent full-width-at-half-maximum (FWHM) of the A exciton peak is plotted in Fig. 1(d). The narrow width of 15 meV at low temperature attests to the high quality of the crystal.[26]

Flakes are mechanically exfoliated directly or transferred to prefabricated backgate structures using a PMMA/PVA stamp or a van der Waals transfer method.[27,28] Four types of backgate stacks are used. These are respectively SiO$_2$/doped Si, h-BN/graphite, HfO$_2$/Au and h-BN/HfO$_2$/Au. The gating efficiency varies from 7×10$^{10}$ /cm$^2$/V to 3×10$^{12}$ /cm$^2$/V (See S2, Ref. 51). We have experimented with the encapsulation of the device using PMMA, h-BN or none. Devices encapsulated by PMMA or h-BN are measured in ambient conditions. Uncapped devices are measured in vacuum. As we will show in Fig. 4, neither the substrate nor the encapsulation layer has a significant effect on the sub-threshold swing of the devices.

Both two-terminal and van der Pauw measurements are made at room temperature using either a constant current source or a constant voltage bias depending on the impedance of the device. Measurements are performed in the linear transport regime with small biases. This corresponds to a source-drain bias $V_{sd} \lesssim$ 100 mV. Both low-frequency lock-in and dc techniques are employed. The backgate voltage $V_{bg}$ is varied, either continuously or in a pulsed mode illustrated in Fig. 2(b).

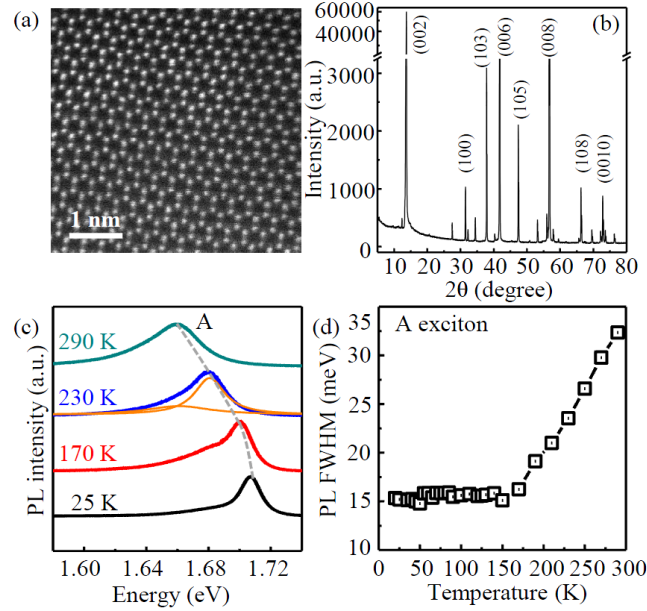

FIG. 1. (Color online) Characterization of synthesized $WSe_2$. (a) High-resolution HAADF TEM image of a $WSe_2$ crystal grown by the first CVT method and imaged along the [110] plane of both W and Se atoms. The hexagonal lattice confirms the 2H phase of $WSe_2$. (b) XRD spectrum of a $WSe_2$ crystal grown by the second CVT method using the $K_\alpha$ line of copper. The Miller indices are indicated in the plot. The lattice constants are $a$=3.28 Å and $c$=12.95 Å, in very good agreement with the literature. (c) The PL spectra of a monolayer $WSe_2$ sheet exfoliated to $SiO_2$ substrate (from crystals grown by method 1) at selected temperatures from 25 to 290 K. The position of the A exciton peak is indicated by the dashed line. Fits to the 230 K trace are shown underneath the data. (d) Temperature dependence of the FWHM of the A-exciton peak. The low-temperature width of 15 meV indicates the high quality of the $WSe_2$ crystal.

Figure 2(a) shows a typical $V_{bg}$-dependent two-terminal conductance $G(V_{bg})$ of a 5L device (5L-A), the optical micrograph of which is shown in the inset. Forward and backward sweeps are shifted from one another by approximately $\Delta V_{bg}$=9.7 V, corresponding to a density difference of $\Delta n$=1.26×10$^{13}$ /cm$^2$. The direction of the hysteresis indicates charge trapping at play. We can suppress this hysteresis completely by applying $V_{bg}$ in pulse in a polarity-alternating sequence illustrated in Fig. 2(b), following methods reported in the literature.[29, 30] (See S3, Ref. 51). The resulting hysteresis-free $G(V_{bg})$ curve is shown in Fig. 2(c). We applied the same method to hysteretic devices in order to remove the contribution from charging the trap states to the sub-threshold swing (SS). We also perform four-terminal van der Pauw or $R_{xx}$ measurements when possible. Figure 2(d) plots the calculated sheet conductance vs the carrier density $\sigma_s(n)$, where $n$ is calculated using the charge neutrality points estimated in Fig. 2(a). On the hole side, $\sigma_s$ reaches a conductance of order $e^2/h$ in the vicinity of $n^*=0.9\times10^{13}$/cm$^2$, which we equate with the mobility edge of the valence band. Among the few-layer (1-5L) WSe$_2$ devices we studied, $n^*$ varies from 0.9-1.2×10$^{13}$/cm$^2$ and the field effect mobility near $n^*$ is typically a few hundred cm$^2$/Vs (~ 300 cm$^2$/Vs in Fig. 2(d)). These values are in good agreement with other $n^*$ and mobilities reported in the literature for high quality devices.[19, 21-23] The large $n^*$ points to a high DoS of localized states inside the band gap of few-layer MX$_2$ materials. In the remainder of the paper, we focus on the conduction *below* the mobility edge, i.e. the subgap regime. We show that gate-modulated thermionic field emission through the Schottky barrier contacts dominates the conductance in this regime and its modeling enables us to determine the DoS of subgap states self-consistently.

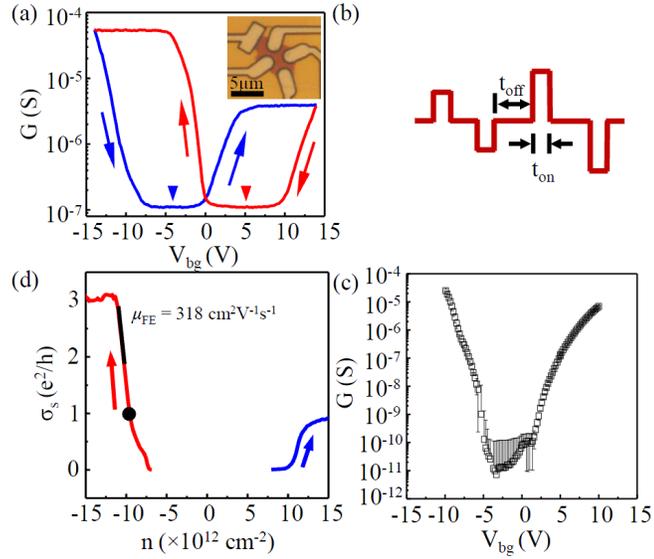

FIG. 2. (Color online) Transport characteristics of a 5L WSe$_2$ transistor (device 5L-A). (a) Two-terminal conductance $G$ vs the backgate voltage $V_{bg}$ from continuous $V_{bg}$ sweep. Arrows indicate the sweeping direction of $V_{bg}$. Triangles mark the charge neutrality points on each sweep. $G(V_{bg})$ flattens at large $V_{bg}$ due to the onset of charge trap screening. The inset shows an optical micrograph of the device. (b) A schematic $V_{bg}$ pulse sequence. $t_{on}$= 25 ms. $t_{off}$=125 ms. (c) $G(V_{bg})$ of the same device obtained in pulsed $V_{bg}$ sweep showing complete suppression of hysteresis. (d) The sheet conductance vs carrier density $\sigma_s(n)$ obtained via the van der Pauw method. Arrows mark the sweep direction during which the data is taken. $n$ is calculated using the charge neutrality point voltages estimated in (a). On the hole side, the mobility edge (black dot) occurs at roughly $n^*=0.9\times10^{13}$/cm$^2$. The field effect mobility $\mu_{FE}$=318 cm$^2$V$^{-1}$s$^{-1}$ for the marked range.

In the subgap regime where both two-terminal and four-terminal measurements are possible, we find the contact resistance $R_c$ dominates the channel resistance $R_{ch}$, i.e. $R_c \gg R_{ch}$. A detailed comparison of $R_c$ and $R_{ch}$ on device 3L-D is given in S5 of Ref. 51. The same conclusion was reached by Liu et al., who systematically studied both using a transfer length method and found $R_c$ to increase more rapidly than $R_{ch}$ as the Fermi level $E_F$ approaches the mid gap.[13] In the following analysis, we assume that $R_c >> R_{ch}$ holds true in the deep subgap regime, where four-terminal measurements become impossible. This assumption is self-consistently justified following the analysis of Fig. 3.

Figure 3(a) plots $G(V_{bg})$ of a 3L device on h-BN/graphite gate stack (3L-B), where the trap-free h-BN/WSe$_2$ interface leads to no hysteresis in continuous $V_{bg}$ sweeps. The symmetry between electrons and holes suggests the work function of Ti/Au contacts roughly aligns with the middle of the band gap $E_i$, as illustrated in the inset. We observe the same phenomenon in Pd (device 3L-D) and graphite (device 4L-E) contacted devices, as shown in Fig. 4, despite the large work function difference between Ti, Pd and graphite. This suggests Fermi level pinning close to $E_i$, presumably by defect states of the WSe$_2$.[18] In the literature, the work function of a variety of contact metals was found to all lie close to the conduction band in MoS$_2$ transistors, presumably due to Fermi level pinning as well.[14] We approximately locate the charge neutrality point, where $E_F=E_i$, by extrapolating $G(V_{bg})$ of both carriers to the intersection of $V^0_{bg}=-0.58$ V and $G_0=8.7\times10^{-14}$ S. Here, the contact resistance is dominated by thermionic emission (TE) over the barrier $\Phi_B=\Phi_{bn}=\Phi_{bp}=1/2\, E_g$. The two-dimensional current density $J$ is given by

$$J = A^*_{2D} T^{3/2} exp\left(-\frac{e\phi_B}{k_B T}\right) \times \left[exp\left(\frac{eV_{sd}}{k_B T}\right) - 1\right], \quad (1)$$

where $A^*_{2D} = \frac{(8\pi k_B^3 m^*)^{1/2} e}{h^2}$ is the two-dimensional Richardson constant.[31] Using $m^*=0.5\, m_0$,[32-34] IV data in the small $V_{sd}$ regime and device dimensions, we obtain an estimate of $\Phi_B=0.69$ eV and $E_g=1.38$ eV. This result agrees very well with the PL emission energy of 1.45 eV observed for our 3-layer WSe$_2$ (See S4, Ref. 51) and in Ref. 35.

The application of a positive $V_{bg}$ moves $E_F$ towards the conduction band edge $E_c$, creating band bending near the contacts as illustrated in Fig. 3(b). The change of the Fermi level

$$\Delta E_F = E_F - E_i = \frac{eC_{bg}}{C_{bg}+C_q}\left(V_{bg} - V^0_{bg}\right), \quad (2)$$

where $C_q = \rho(E)e^2$ is the quantum capacitance of the sheet per area and $\rho(E)$ the DoS of the impurity states inside the band gap of WSe$_2$. Equation (2) does *not* include the contribution of the charge trap states, since they are either absent (in h-BN/graphite devices) or are not activated in the pulsed gate measurements shown in Fig. 2(c). Fast trap states with response time less than a few ms have densities $\lesssim 1\times10^{12}$/cm$^2$ for typical oxides,[36] which is an order of magnitude smaller than $C_q$ values extracted below. Equation (2) has two limits. In the limit of $C_q << C_{bg}$, which can be realized in very clean samples or using electrolyte gating,[37] $\Delta E_F = e\Delta V_{bg}$, i.e. the movement of $E_F$ follows that of the gate voltage. In the opposite limit of $C_q \gg C_{bg}$, which corresponds to a large number of impurity states inside the band gap, moving $E_F$ through the band gap $E_g$ requires a large gate voltage range $e\Delta V_{bg} = \left(\frac{C_q}{C_{bg}}\right) \times E_g$. The presence of the impurity states, however, reduces the depletion width of the Schottky barrier $x_{dep}$ and promotes quantum tunneling through the Schottky barrier, i.e. field emissions (FE) and thermionic field emissions (TFE), in addition to thermionic emissions (TE) over the barrier.[38-40] As illustrated in Fig. 3(b), the TFE mechanism combines thermal excitation and quantum tunneling. Its 2D current density $J$ (in the small $V_{sd}$ limit) can be adapted from Equations (88-92) of Chapter 3 of Ref. 41 and reads:

$$J_{TFE} = \frac{A_{2D}^{**} T^{1/2} \sqrt{\pi E_{00} E_b}}{k_B \cosh(E_{00}/k_B T)} exp\left[-\frac{E_c - E_F}{k_B T}\right] exp\left[-\frac{E_b}{E_0}\right], \quad (3)$$

where, $E_b = \Delta E_F$ is the band bending shown in Fig. 3(b).

Here, $E_0 = E_{00} \coth(E_{00}/k_B T)$, and $E_{00} = \frac{e\hbar}{2}\sqrt{\frac{N_i}{m^* \varepsilon}}, \quad (4)$

where $N_i$ is the impurity density of the material in units of cm$^{-3}$ and $\varepsilon$ the dielectric constant. $m^*$ is the effective mass.

The two exponential terms of Equation (3) capture the two key ingredients of the TFE process, i.e. thermal activation to the conduction band edge and the tunneling process characterized by $exp\left[-\frac{E_b}{E_0}\right]$. $E_0$ and $E_{00}$ are important energy scales of the problem. A large $N_i$ leads to large $E_{00}$ and $E_0$, which enhance the tunneling probability. Tunneling at the band edge, i.e. field emission, occurs when $E_{00} \gg k_B T$, e.g. in heavily doped semiconductors or at low temperature. When $E_{00} \ll k_B T$, carriers need to be thermally excited over the barrier (TE). TFE occurs in between the two limits, where tunneling occurs somewhere along the barrier as illustrated in Fig. 3(b).

Equations (2) and (3) together lead to the expression for the sub-threshold swing $SS \equiv \left[\frac{d\log J}{dV_{bg}}\right]^{-1}$ given in Equation (5):

$$SS = \left(\frac{E_0}{E_0 - k_B T}\right)\left(1 + \frac{C_q}{C_{bg}}\right) \times \frac{k_B T}{e} ln10 \text{ /decade}. \quad (5)$$

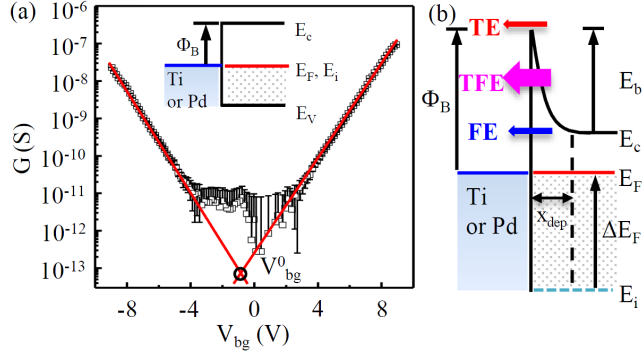

FIG. 3. (Color online) Transmission through a Schottky barrier contact. (a) Two-terminal conductance $G$ vs $V_{bg}$ for a 3-layer device on h-BN (device 3L-B) in a semi-log plot. The absence of hysteresis indicates trap-free interface. Fits to log $G$ vs $V_{bg}$ yield SS of 1.6 V/decade for both electron and hole. The charge neutrality point occurs at $V_{bg}^0$= -0.58 V and $G_0$=8.7×10$^{-14}$ S, the band diagram at which is shown in the inset. (b) Band diagram near the metal contact in the case of electron doping. $E_b=\Delta E_F$. Three transmission mechanisms are illustrated. TE represents thermal excitation over the Schottky barrier. FE represents direct tunneling at the band edge. TFE combines thermal excitation and tunneling at intermediate barrier height.

Here, $k_BT$=26 meV and we neglect the weak $V_{bg}$ dependence of the prefactors in Equation (3). The 2D impurity density in a thin WSe$_2$ sheet is given by $N_i t$, where $t$ is the thickness of the sheet. Assuming each impurity provides ~ one subgap state, $N_i t$ is approximately the same as the total number of subgap states, i.e.

$$N_i t = \rho(E)E_g = C_q E_g/e^2. \qquad (6)$$

Here, we treat $\rho(E)$ and $C_q$ as average quantities and replace integration with simple multiplication. $\rho(E)$ does appear to be approximately constant for a large range of subgap energy in our devices, as revealed by the linear log$G$–$V_{bg}$ relation in Figs. 3 an 4. Equations (4)-(6) together allow us to self-consistently estimate microscopic parameters $N_i$ and $\rho(E)$ using the measured SS. We use $m^*$=0.5 $m_0$ and $\varepsilon$=4.63 in our calculations.[32-34] For example, device 3L-B shown in Fig. 3(a) exhibits SS=1.6 V/decade for both electrons and holes. Using $t$=2 nm and $E_g$=1.45 eV, we obtain $\rho(E)$=1.6×10$^{13}$ /cm$^2$/eV and $N_i$=1.2×10$^{20}$ /cm$^3$. The calculated $E_{00}$=130 meV=5 $k_BT$ at room temperature, thus validating the applicability of the TFE regime. $\rho(E)$=1.6×10$^{13}$ /cm$^2$/eV also predicts a mobility edge carrier density of $n^* = \rho(E)E_g/2$=1.2×10$^{13}$ /cm$^2$, consistent with the observed values.

Similar analyses are performed on ten few-layer (1-5L) devices exfoliated from WSe$_2$ crystals synthesized using the two recipes described in S1 of Ref. 51. Overall, we find $N_i$ to be in the range of 0.3-1.3×10$^{20}$ /cm$^3$ and $E_{00}$ in the range of 3-5 $k_BT$. The subgap localized DoS $\rho(E)$ ~ 1-2×10$^{13}$ /cm$^2$/eV. Such large $N_i$ is equivalent to heavy doping in conventional semiconductors, where TFE and FE transmissions were found to occur at room temperature.[39-41] The large $N_i$ will also lead to substantial hopping conduction through the localized states in the WSe$_2$ channel. Since $\rho(E)$ is roughly a constant for a large range of subgap energies, this hopping conductivity maintains at a relatively high level. In contrast, the transmission through the Schottky barrier contacts exponentially decays as $E_F$ moves towards mid gap. The different energy dependence provides a self-consistent justification of $R_c \gg R_{ch}$ in the subgap regime and explains the observations of ours and that of Liu *et al.*[13]

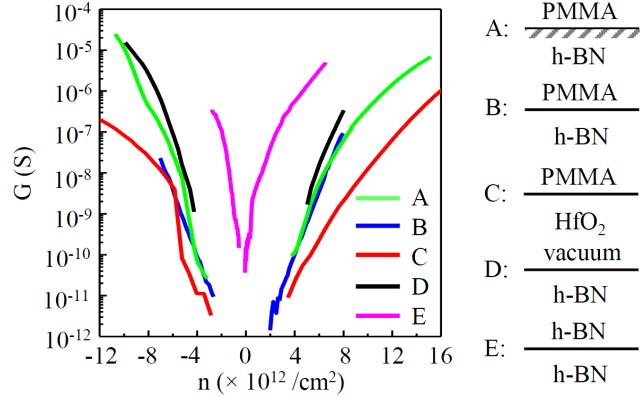

FIG. 4. (Color online) Comparison of devices in different dielectric environment. Two-terminal conductance $G$ vs carrier density $n$ for devices 5L-A, 3L-B, 5L-C, 3L-D and 4L-E. Schematics indicate the dielectric layers adjacent to the WSe$_2$ sheet. The complete gate stacks are h-BN/HfO$_2$/Au, h-BN/graphite, HfO$_2$/Au, h-BN/graphite, h-BN/SiO$_2$/Si and the gating efficiencies are 1.3, 0.84, 3.0, 0.61, and 0.06×10$^{12}$/cm$^2$/V respectively for devices A to E. After accounting for the gating efficiencies, the SS slopes are very similar over a large range of subgap energies despite the large difference in substrate surface chemistry.

The above analyses make it clear that the gate modulation of the two-terminal conductance in our few-layer WSe$_2$ transistors is primarily achieved by controlling the transmission through the Schottky barrier contacts. This type of behavior, i.e. a Schottky barrier transistor, was also found in semiconducting carbon nanotubes.[42] Furthermore, the transmission through the contact barrier is a combination of thermal excitation and tunneling, due to a large number of states existing inside the band gap that lead to reduced barrier width near the contacts.

We have fabricated WSe$_2$ devices embedded in a variety of dielectric environment/encapsulation (combination of vacuum, PMMA, h-BN, SiO$_2$ and HfO$_2$) to shed light on the origin of the subgap states. Overall, we have not found any systematic dependence of $\rho(E)$ on the choice of the environment. As an example, Fig. 4 compares $G$ ($n$) of five devices embedded in different environment. All five exhibit similar SS slopes in the subgap regime while the chemistry and dielectric constant of the environment differ greatly. The SS remains large even in devices encapsulated by clean h-BN (Device 4L-E). This indicates that at the level of 1×10$^{13}$/cm$^2$/eV, the subgap states are dominated by internal contributions rather than interface states that are known to exist in oxides. This is consistent given that oxide charge traps are typically on the order of $10^{11}$-$10^{12}$ /cm$^2$,[36, 43, 44] which is too small to account for the $\rho(E)$ observed here. It should also be emphasized that scenarios explored here pertain to the range of $E_F$ not too close to the band edge. As Fig. 4 shows, $G(n)$ curves as $E_F$ approaches $E_c$ or $E_v$, suggesting the appearance of additional impurity states. Substrate-related impurity states are primary candidates.[45, 46] In addition, the assumption of $R_c \gg R_{ch}$ may not hold anymore as $E_F$ approaches $E_c$ or $E_v$ and the contacts become transparent. The analysis of this regime thus requires the separation of the two, via four-terminal measurements for example.

Recent experiments and simulations have shown that a rich variety of structural defects, such as chalcogen vacancies and dislocations at grain boundaries, can create defect states with a wide span of subgap energies.[47-50] Defect density on the order of 1%, such as that observed in STM studies of MoS$_2$[50] can potentially account for the phenomena observed here. Such low density of defects is difficult to assess using conventional microscopy and elemental analysis but has a high impact on electronic properties. In addition, few-layer TMDC devices are vulnerable against the degradation caused by interactions with the environment (e.g., oxygen, humidity), which may also play a role in creating additional impurity states.

In summary, we studied the electrical transport properties of few-layer WSe$_2$ transistors in the subgap regime. We demonstrate that the gate modulation of the two-terminal conductance originates from controlling the thermionic field transmission through the Schottky contact barrier. Underlying such behavior is a large number of localized states inside the band gap of the material. Further understanding and elimination of these impurity states will prove essential towards improving the qualities of TMDC materials and devices, thus opening the door to the exploration of fundamental phenomena in these fascinating 2D systems.

We are grateful for helpful discussions with Suman Datta. J.W., S.F., T.E.M., M.T. and J.Z. are partially supported by NSF MRSEC Grant No. DMR-0820404. M.A.T.N., D.R., T.E.M., L.B. and M.T. are supported by the U.S. Army Research Office MURI Grant No. W911NF-11-1-0362. K.W. and T.T. are supported by the Elemental Strategy Initiative conducted by the MEXT, Japan. T.T. is also supported by a Grant-in-Aid for Scientific Research on Grant 262480621 and on Innovative Areas "Nano Informatics" (Grant 25106006) from JSPS. Part of this work is performed at The NHMFL, which is supported by NSF through NSF-DMR-0084173 and the State of Florida. The authors acknowledge use of facilities at the PSU site of NSF NNIN.

# Gate-modulated conductance of few-layer WSe$_2$ field-effect transistors in the subgap regime: Schottky barrier transistor and subgap impurity states (Supplementary material)

Junjie Wang[1], Daniel Rhodes[2], Simin Feng[1], Minh An T. Nguyen[3], K. Watanabe[4], T. Taniguchi[4], Thomas E. Mallouk [1,3,5], Mauricio Terrones[1,3,6,7], Luis Balicas[2], J. Zhu[1,7]

[1]Department of Physics, The Pennsylvania State University, University Park, PA, 16802, USA
[2]National High Magnetic Field Lab, Florida State University, FL, 32310, USA
[3]Department of Chemistry, The Pennsylvania State University, University Park, PA, 16802, USA
[4]National Institute for Materials Science, 1-1 Namiki, Tsukuba, 305-0044, Japan
[5]Department of Biochemistry and Molecular Biology, The Pennsylvania State University, University Park, PA, 16802, USA
[6]Department of Materials Science and Engineering, The Pennsylvania State University, University Park, PA, 16802, USA
[7]Center for 2-Dimensional and Layered Materials, The Pennsylvania State University, University Park, PA 16802, USA

**S1. WSe$_2$ bulk crystal synthesis**

We employ two procedures to synthesize WSe$_2$ crystals. The first technique uses a chemical vapor transport technique using either iodine or excess Se as the transport agent. 99.999% pure W powder and 99.999% pure Se pellets were introduced into a quartz tube together with 99.999% pure iodine. The quartz tube was vacuumed, brought to 1150 $^o$C, and held at this temperature for 1.5 weeks at a temperature gradient of < 100 $^o$C. Subsequently, it was cooled to 1050 $^o$C at a rate of 10 $^o$C per hour, followed by another cool down to 800 $^o$C at a rate of 2 $^o$C per hour. It was held at 800 $^o$C for 2 days and subsequently quenched in air.

The second technique first synthesizes WSe$_2$ powder by heating a mixture containing stoichiometric amounts of tungsten (Acros Organics 99.9%) and selenium (Acros Organics 99.5+%) together at 1000$^o$C for 3 days in an evacuated and sealed quartz ampoule (10 mm ID, 12 mm OD, 150 mm length).  The mixture was slowly heated from room temperature to 1000$^o$C for 12 hours to avoid any explosion due to the strong exothermic reaction.  Chemical vapor transport was then used to grow WSe$_2$ crystal from the powder using iodine (Sigma-Aldrich, 99.8+%) as the transport gas at 2.7 mg/cm$^3$ for 10 days in an evacuated and sealed quartz ampoule (10 mm ID, 12 mm OD, 127 mm length).  The source and growth zones were kept at 995$^o$C and 851$^o$C, respectively.  The resulted crystals was washed with hexane and dried in vacuum to remove any residual iodine and hexane. WSe$_2$ powder and crystals were analyzed using XRD, ESEM, ESEM-EDS.  XRD confirm that both the powder and crystal WSe$_2$ synthesized were pure 2H phase.

## S2. Gate stack and device fabrication

Four types of backgate stacks are used. The 300 nm SiO$_2$/Si substrate is used as purchased and has a gating efficiency of $7\times10^{10}$ cm$^{-2}$. The h-BN/graphite gate is made by first mechanically exfoliating uniform multi-layer graphite sheets (3-5nm) to a SiO$_2$/Si substrate and then transfer h-BN flakes of ~ 20nm thickness using a PMMA/PVA stamp[1]. The assembly is then rinsed in acetone and annealed in a Ar (90%) /H$_2$ (10%) mixture at 450°C for 3 hours to remove PMMA residues. The h-BN/graphite gate provides a clean, trap-free h-BN/WSe$_2$ interface with a gating efficiency close to $1\times10^{12}$ /cm$^2$ depending on the thickness of the h-BN. The HfO$_2$/Au gate is fabricated by RIE etching of a SiO$_2$/Si substrate followed by metal deposition and ALD growth of ~ 40 nm HfO$_2$. The details of the lithography is given in the supporting information of Ref.[2] and the ALD recipe can be found in the supporting information of Ref.[3]. The HfO$_2$/Au gate has a large gating efficiency of $3\times10^{12}$ /cm$^2$ and can achieve carrier density beyond $3\times10^{13}$ /cm$^2$. We have also tried to combine the large gating efficiency of the HfO$_2$/Au stack and the cleanness of the h-BN substrate by transferring an h-BN sheet onto the HfO$_2$/Au stack. Here a polydimethylsiloxane (PDMS) stamp is used to transfer the h-BN sheet following Ref[4]. No water is involved and no annealing is done after peeling off the PDMS stamp. WSe$_2$ flakes are transferred directly to the h-BN/HfO$_2$/Au stack. We find the gate-dependent conductance $G(V_{bg})$ to be hysteresis-free in a small range of $V_{bg}$ but exhibits strong saturation at large $V_{bg}$'s, suggesting the activation of charge trap states that completely screens the WSe$_2$ channel (See Figure 2(a) in the text). AFM AC mode scanning of the PDMS-transferred h-BN surface reveals a thin layer of PDMS residue with surface roughness ranging from 0.3nm to 1.5nm. We attribute the charge traps to the presence of the PDMS residue. The h-BN/HfO$_2$/Au gate has a gating efficiency of $1.3\times10^{12}$ /cm$^2$.

Multi-terminal devices are fabricated by exfoliating or transferring (PMMA/PVA stamp) few-layer WSe$_2$ sheets to the gate stacks. We also use the dry van der Waals transfer technique[5] to sequentially pick up h-BN, graphite electrodes, WSe$_2$ and h-BN and deposit the complete stack to a SiO$_2$/Si backgate to make graphite-contacted, h-BN encapsulated devices. E-beam lithography and metal deposition are used to make Ti (5nm)/Au (40-50nm) and Pd (45 nm) contacts to WSe$_2$. Two e-beam doses, 330μC/cm$^2$ and 510 μC/cm$^2$, were used. The 330μC/cm$^2$ dose, together with a MIBK:IPA 1:1 developer, is effective in clearing exposed PMMA on graphene, BN and SiO$_2$ but leaves a thin resist residue layer of 0.6 to 0.9 nm thick on dichalcogenide materials (Figure S1(a)). A gentle Ar$^+$ bombardment step was done *in situ* in a Kurt J. Lesker Lab-18 system prior to the metal deposition. A discharge voltage of V=75V, an emission current of I$_e$=0.4A, and a duration of t=2s can completely remove the resist residue as Figure S1(b) shows. Alternatively, we have used a high dose of 510 μC/cm$^2$ to completely clear the exposed resist prior to metal deposition.

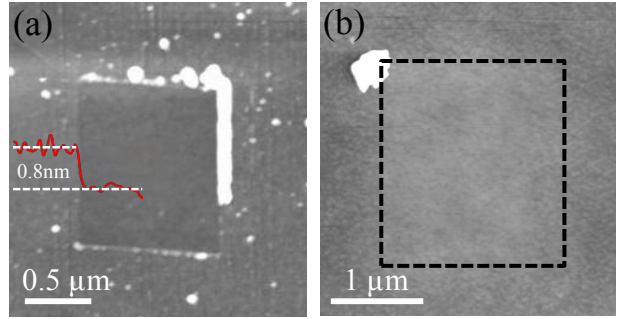

Figure S1. AFM study of a monolayer $MoS_2$ after e-beam writing and developing. (a) AFM tapping mode image shows resist accumulation at the borders of a square area scanned by the tip in contact mode prior to the acquisition of this image. The line cut reveals a resist layer thickness of 0.8 nm. (b) Another AFM tapping mode image shows a clean surface after low-energy $Ar^+$ bombardment. The A-exciton of the sheet is reduced by about a factor of 2 after $Ar^+$ bombardment. We did not observe any appreciable difference between devices made this way and those that used a high e-beam dose only.

## S3. Measurement of $\sigma(V_{bg})$ in pulsed $V_{bg}$ mode

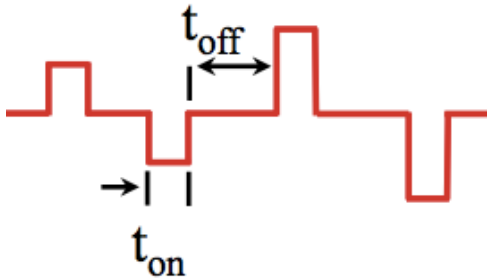

Figure S2: A $V_{bg}$ pulse sequence. $t_{on} = 25$ ms and $t_{off} = 125$ ms.

We apply $V_{bg}$ in pulse to eliminate the hysteresis of $G(V_{bg})$ in some devices. An alternating sequence of positive and negative $V_{bg}$'s is used. Each $V_{bg}$ is on for $t_{on} = 25$ ms, followed by $V_{bg} = 0$ for a duration of $t_{off} = 125$ ms. The two-terminal conductance is obtained by sourcing a small DC voltage $V_{sd}$ (10 to 100 mV) and measuring the current using a current preamp. Both positive and negative $V_{sd}$ are used and the results averaged to remove any offset in $V$ and $I$. Data after the equilibrium is reached after each pulse is applied are used.

## S4. Photoluminescence of few-layer WSe$_2$

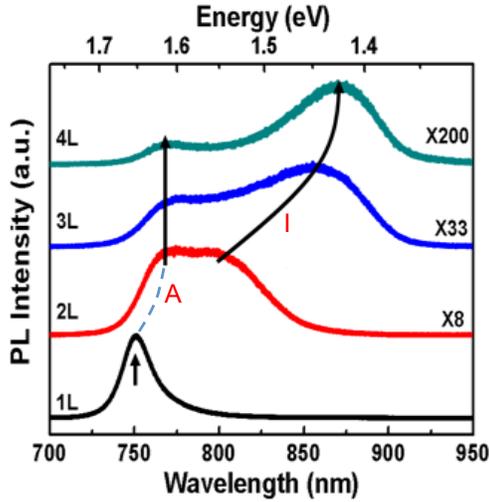

Figure S3: Photoluminescence spectra of 1-4 layer WSe$_2$ showing the direct gap emission (A peak) in monolayer and indirect gap emissions (I peak) in few-layers. The emission energies are used to approximate the band gap of WSe$_2$. The excitation laser line is 488 nm.

## S5. Comparison of two-terminal resistance, contact resistance, and WSe$_2$ channel resistance

Figure S4 plots the $V_{bg}$ dependence of the two terminal resistance $R_{2t}$, the contact resistance $R_c$ and the resistance of the WSe$_2$ channel $R_{ch}$ in device 3L-D. $R_{ch}$ is estimated by scaling the resistance measured in the four-terminal geometry $R_{xx}$ using $R_{ch}=(L_2/L_1)\times R_{xx}$, where L$_2$ and L$_1$ are indicated on the optical micrograph of the device shown as the inset. Contact resistance $R_c$ is calculated using $R_c=R_{2t}-R_{ch}$. It is clear from the comparison that $R_c \gg R_{ch}$ and $R_{2t} \approx R_c$ throughout the $V_{bg}$ range, where $R_{xx}$ can be measured reliably.

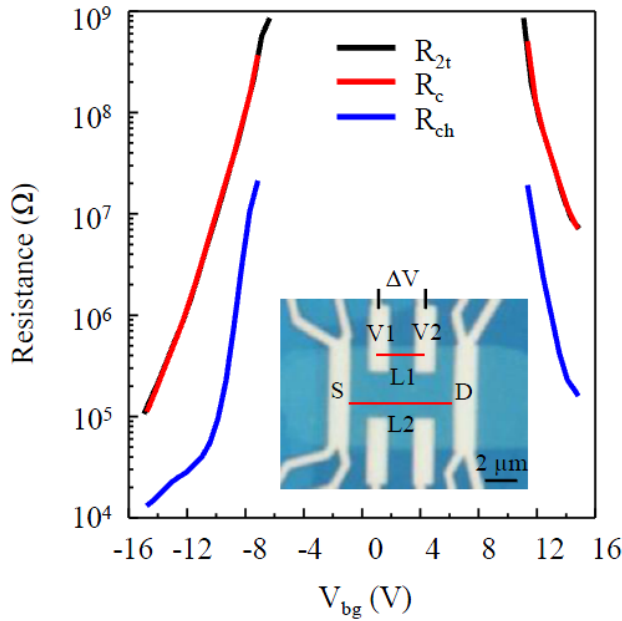

Figure S4. Two-terminal resistance $R_{2t}$ (black line), contact resistance $R_c$ (red line), and channel resistance $R_{ch}$ (blue line) as a function of $V_{bg}$ on a semi-log plot. Inset shows an optical micrograph of the device. $L_2/L_1 = 2.2$.